\documentclass[12pt]{spieman}  
\usepackage{amsmath,amsfonts,amssymb}
\usepackage{graphicx}
\usepackage{setspace}
\usepackage{tocloft}
\usepackage{subfig}
\usepackage{algorithm}
\usepackage{algpseudocode}

\usepackage{multirow}
\usepackage{multicol}
\usepackage[table]{xcolor}
\usepackage[para, online, flushleft]{threeparttable}
\usepackage{booktabs}

\title{Fusion of heterogeneous bands and kernels in hyperspectral image processing}

\author[a]{Muhammad Aminul Islam}
\author[b]{Derek T. Anderson}
\author[c]{John E. Ball}
\author[c]{Nicolas H. Younan}
\affil[a]{University of Missouri, Center for Geospatial Intelligence, Columbia, Missouri, USA, 65211}
\affil[b]{University of Missouri, Department of Electrical Engineering and Computer Science, Columbia, Missouri, USA, 65211}
\affil[c]{Mississippi State University, Department of Electrical and Computer Engineering, Mississippi State, Mississippi, USA, 39762}

\cftpagenumbersoff{figure}
\cftpagenumbersoff{table} 
\begin{document} 

\vspace*{5cm}
Author's copy. Accepted for publication in the Journal of Applied Remote Sensing.
\newpage

\maketitle

\begin{abstract}
Hyperspectral imaging is a powerful technology that is plagued by large dimensionality. Herein, we explore a way to combat that hindrance via non-contiguous and contiguous (simpler to realize sensor) band grouping for dimensionality reduction. Our approach is different in the respect that it is flexible
and it follows a well-studied process of visual clustering in high-dimensional spaces. Specifically, we extend the  \textit{improved
visual assessment of cluster tendency} and \textit{clustering in ordered dissimilarity data} unsupervised clustering algorithms for supervised hyperspectral learning. In addition, we propose a way to extract diverse features via the use of different proximity metrics (ways to measure the similarity between bands) and kernel functions. The discovered features are fused with $l_{\infty}$-norm multiple kernel learning. Experiments are conducted on two benchmark datasets and our results are compared to related work. These datasets indicate that contiguous or not is application specific, but heterogeneous features and kernels usually lead to performance gain. 
\end{abstract}

\keywords{band grouping, feature fusion, $\ell_p$-norm MKL, visual clustering, dimensionality reduction, hyperspectral, VAT, CLODD.}


{\noindent \footnotesize\textbf{*} Corresponding author: Muhammad Aminul Islam,  \linkable{mig5g@missouri.edu} }

\begin{spacing}{2}   

\section{Introduction}
\label{sect:intro}  
Hyperspectral imaging is a demonstrated tool for numerous earth and space-borne applications involving target detection \cite{bajorski2012target,jay2012underwater,zhao2014robust}, invasive species monitoring \cite{somers2013multi,somers2013invasive}, and precision agriculture \cite{hunt2014remote,zhao2015data}. However, the field suffers from the ``curse of dimensionality'' (spatial, spectral and temporal). Of interest is new theory for dimensionality reduction or identification of fewer spectral bands for purposes like multispectral vs hyperspectral imaging, typically relative to some task, which ultimately aids efficient computation, storage, transmission, classification and lower system cost. While numerous noteworthy methods have been explored, effectiveness and efficiency of search, fusion and classification remain unsolved.  


In general, the vast majority of band selection and band grouping algorithms can be divided into three categories---projection, clustering, and search-based. Projection methods construct ``features'' via some non-linear technique like manifold theory \cite{chapel2014perturbo,ziemann2015hyperspectral} or a linear combination of the bands. Well-known approaches include \textit{principal component analysis} (PCA)\cite{farrell2005impact}, \textit{Fishers linear discriminant analysis} (FLDA), \textit{generalized discriminant analysis} (GDA) \cite{bandos2009classification}, \textit{random projections} (RP) \cite{FowlerPaper}, and kernel extensions. Some methods are unsupervised, e.g., PCA and RP, and others are supervised, e.g., FLDA and GDA. Clustering is unsupervised learning and it can be applied to hyperspectral imagery in numerous ways. While it does not directly do dimensionality reduction, it helps identify structure and one can take that information and use it for dimensionality reduction or band (group) selection. For example, in \cite{martinez2007clustering} Martinez et al. used an information measure to compute dissimilarity between bands and hierarchical clustering with Ward's single linkage to produce a minimum variance partitioning of the bands. In \cite{imani2014band}, Imani and Ghassemain used (hard) $c$-means. 
In \cite{ball2014proximity}, Ball et. al. explores the notion of grouping contiguous similar bands from a proximity matrix, which 
computes the mean value of a block in a proximity matrix and compares it with a threshold to determine the edge between clusters.   Herein, we refer to this method as BG-Mean.
However, these clustering approaches are limited due to factors like lack of sophistication of clustering algorithm and how they are used, i.e., dendograms and metric and threshold selection and initialization, selection of $c$, and ability to address outliers in $c$-means. 


After selecting a suitable bandgrouping method for a particular task, the performance of an application can significantly be improved by employing different techniques to extract and fuse diverse sets of features.
\textit{Multiple kernel learning} (MKL) is one such technique for feature selection and feature space fusion and has extensively been used in applications with objective of target detection or classification.
MKL learns the weights for optimum combination of the kernels that project the  low dimensional features in the high dimensional Reproducing Hilbert Kernel Space (RHKS) where they are linearly separable. Depending on how the weights are regularized, MKL's variant is termed as $\ell_p$-norm or p-norm MKL. Since the MKL is computationally very expensive, a lot of research efforts have been put forth for developing efficient algorithm, which results in numerous MKL algorithms. They work on the same problem but differ in the way it is solved and their efficiency depends on different aspect of the problem, such as some are very good with large number of kernels while some others work well with large scale data-set. Some algorithms runs efficiently on $\ell_1$-MKL and some takes very few iterations to converge or has a closed form solution. Among MKL variants, the $\ell_{\infty}$-norm MKL, alternatively called as the unweighted-sum SVM kernel, is unique in the sense that the weights are all fixed ($1$s) and therefore can be solved with \textit{support vector machine} (SVM) with the sum of kernels instead of a single kernel. Since there is no weights to learn, $\ell_{\infty}$-norm MKL is computationally most efficient among MKLs.  Moreover, dense fusion (higher norm including $\ell_{\infty}$-norm MKL) is better for diverse and complementary features \cite{bleakley2007supervised} while $\ell_1$-norm is good for noisy kernels. 
Bleakley et. al. explored the idea of directly summing all kernels instead of learning their weights \cite{bleakley2007supervised}, which showed to have the best results in compared to variably weighted $\ell_p$-norm MKLs. 
A detail analysis and comparison of the algorithms can be found in Gonen and Alpaydın \cite{gonen2011multiple}.

In hyperspectral image processing, MKL has been used for different purposes such as classification \cite{gu2012representative} and image feature selection \cite{tuia2010learning}. In \cite{gu2012representative}, a computationally efficient formulation of MKL termed as \textit{representative MKL} (RMKL) is proposed for performing  classification with high dimensional hyperspectral data set. A number of RBF kernel is created by varying the width parameter of the RBF kernel function. Then it finds the linear combination of the kernels that yields max-variance kernel with minimum F-norm error. In \cite{tuia2010learning}, the authors use simpleMKL \cite{alain2008simple} for learning relevant image feature and proposes an algorithm to optimize the width parameter of RBF kernel. The kernels are applied to single feature, group of features and features from heterogeneous sources. Then MKL is used to select the features as well as to fuse them in the feature space. All these papers use only single type of kernels, specifically standard RBF kernel and vary the width parameters of the kernel function to get a set of kernels. While \cite{islam2016fusion} explored two different kernels, rbf and correlation, they did not apply both types of kernels to the same set of features. Islam et. al. also introduced higher $\ell_p$-norm MKL in hyperspectral imagery and showed its power when combining diverse features relative to lower-norm MKL, however, their analysis did not go beyond 100-norm.


Herein, we propose an end-to-end method for feature extraction and fusion. The feature extraction step involves the followings: choose a proximity measure, than use a dimensionality reduction approach based on visual clustering based band grouping, and finally apply kernels to generate diverse features sets in RHKS space. Fusion steps involve directly summing up the kernels, an efficient way of aggregating diverse sets of features in MKL, which does not require learning the weight of each kernel.

We explore a comprehensive clustering approach to grouping and selection of contiguous or non-contiguous bands. Most are aware of clustering algorithms such as (crisp) $c$ (aka $k$) means, i.e., given a set of data $X$, and set of assumptions on the patterns, e.g., desire to partition compact and well separated clouds, find the cluster prototypes and associated parameters. Furthermore, many are aware of the field of (internal or external) cluster validation, e.g., for different selection of clustering algorithm parameters, initial conditions, etc., what is the ``best'' selection? However, few are aware of a third and initial step in clustering called \emph{cluster tendency assessment} (CTA). CTA tries to address fundamental questions like, is there any evidence that there are any clusters at all, if so, what ``type'' of clusters are they, what distance measure or clustering algorithm should we use, etc. 
In this paper, we address all three areas of clustering. First, CTA is engaged via a proximity measure, which produces a $b \times b$ \textit{dissimilarity matrix} (DM) for $b$ bands. In traditional data clustering, the DM is $N \times N$ for $N$ objects. The DM is a rich source of information that has been used by experts and (less sophisticated) band group selection algorithms \cite{ball2014proximity} with success. The challenge is how to define and find structure in a DM. Typically, structure takes the form of blocks of high contrast on the diagonal. If the user desires non-contiguous band groups, then we order the bands (objects in traditional clustering) via the \textit{visual assessment of cluster tendency} (VAT) \cite{bezdek2007visual} or \textit{improved VAT} (iVAT) \cite{havens2012efficient}, where iVAT provides contrast enhancement. If a user desires to automatically find clusters (band groups), then the \textit{clustering on ordered dissimilarity data} (CLODD) \cite{havens2009clustering} algorithm is used (the clustering algorithm step). A feature of CLODD-based clustering is it provides a way to reject bands (objects in standard clustering) that are of little utility (e.g., outliers). However, finding an ``answer'' from CLODD is based on two parameters (discussed later in the article). While it is often possible to pick general parameters relative to some reasonable relative contrast amount and minimum number of bands in a band group for a task like unsupervised band grouping, in supervised band grouping we conduct a parameter search relative to a more concrete criteria, classification accuracy based on the resultant band groups. This operation \emph{closes the loop} and acts like cluster validation. In particular, it is performed because we observed that in some cases what a human ``sees'' in the DM and what a classifier prefers is not necessarily the same. Additionally, it is not always clear where band group boundaries reside.  

This method enable generating diverse features through the use of different proximity measures and kernels. 
Particularly, we use two proximity features (e.g., square of Euclidean and correlation) for CLODD based band grouping and feature extraction. The square of Euclidean measures the distance between points in a multidimensional whereas correlation measures the degree of similarity between patterns and does not depend on distance. Therefore, having distinct characteristics of the proximity measures, we are capturing different aspect of the data via its features. 
While numerous proximity measures, such as Euclidean, correlation, Jeffrey K. Matusita, \textit{spectral angle mapper} (SAM), etc.  have been used in hyperspectral image processing; the use of proximity measures in kernel for MKL or \textit{support vector machine} (SVM) has been  limited by the requirement that the kernel has to satisfy the Mercer's kernel properties. Having that restriction, not all proximity measures can be used in a kernel. Herein, we have used square of euclidean and correlation based RBF for kernels, which have been proven to satisfy the Mercer's theorem.


In summary, we put forth an \emph{end-to-end} clustering approach for band group selection that has roots in visual structure identification in a hyperspectral DM (VAT/iVAT) and we extend CLODD to not rely on user parameters but classification performance. Furthermore, our method has the advantage that we can allow contiguous or non-contiguous band groups. In \cite{CLODDBGslection} we conducted a preliminary small scale investigation into this topic. In the current paper, we outline a search procedure for parameter selection in supervised learning, more data sets are processed to better understand the performance of the proposed ideas and a more in depth analysis into how many final band groups we obtain is performed. Second, we also present how to generate diverse features using different proximity measures and kernels functions. Finally, we apply $\infty$-norm MKL for feature space fusion that is computationally efficient and has superior performance than SVM, commonly used $\ell_1$ or $\ell_2$-norm MKL or little used $\ell_p$-norm MKL \cite{gu2012representative, tuia2010learning}.
\begin{figure*}[!t]
  \centering
\includegraphics[width=0.89\textwidth]{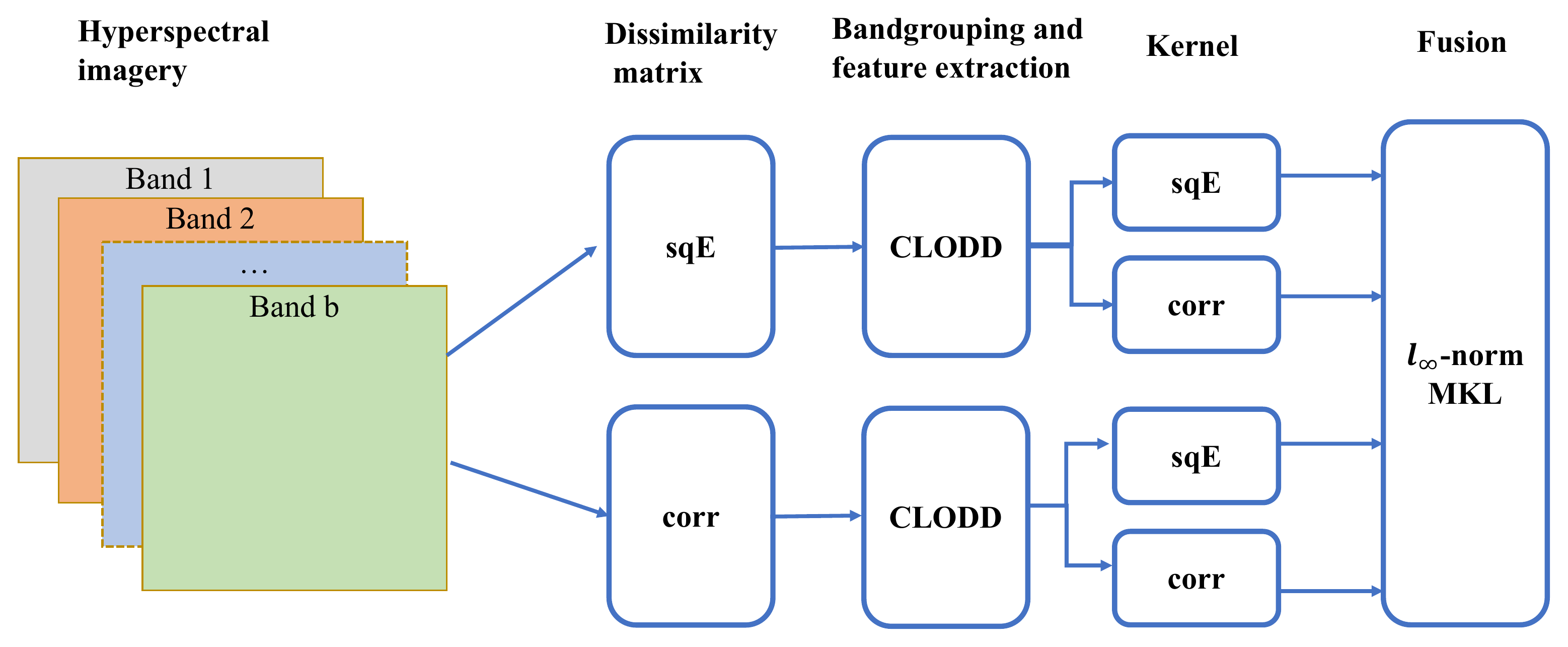}
\caption{Block diagram for the proposed method. In this diagram, sqE stands for the square of Euclidean and corr for correlation.}
\label{fig:block_diagram}
\end{figure*} 

\section{Methods}
\label{sec:methods}
For ease of computation, we rearrange the 3D hyperspectral data cube to form a 2D data set, since no spatial information is incorporated for band grouping; therefore, no information with respect to spectral bands will be lost. 
Each row in this 2D data set now represents a pixel in the image and each column is a band. Let the data set be $ \mathbf{X} = \left\{\mathbf{x}_1, \mathbf{x}_2, \cdots, \mathbf{x}_n \right\}  \in \mathbb{R}^{n \times b} $, where $n$ is the number of pixels in the image and $b$ is the number of bands. The label for each pixel, $\mathbf{x}_i$ is $y_i \in \left\{1, 2, \cdots, L \right\}$, where $L$ is the number of classes. Figure \ref{fig:block_diagram} shows the major steps in the proposed approach.

\subsection{Dissimilarity matrix}
There are numerous proximity measures and their combinations that can, and have, been used. For example, popular unsupervised proximity measures are correlation, Bhattacharyya distance and Kullback-Liebler divergence \cite{ball2014proximity}. 
In this paper, we used two dissimilarity measures based on the square of Euclidean and correlation, the definitions of which are as follows. 
\paragraph{The square of Euclidean:}It is computed as the square of the $l_2$ norm of the differences between pixel values for those two bands, $D(i,j) = || X_{.,i} - X_{.,j} ||^2$, where $X_{.,i}$ is a column vector of all pixel values for band $i$. 

\paragraph{Correlation:} Pearson's correlation coefficient a.k.a. correlation measures the similarity between patterns of two bands.
\[S(i,j) = \frac{COV(X_{.,i}, X_{.,j})}{\sigma_{X_{.,i}} \sigma_{X_{.,j}}},\]
where $\sigma_{X_{.,i}}$ and $\sigma_{X_{.,j}}$ are the variances of bands $i$ and $j$ respectively and $COV(X_{.,i}, X_{.,j})$ is the covariance between bands $i$ and $j$ calculated as $ COV(X_{.,i}, X_{.,j}) = E[(X_{.,i} - E(X_{.,i})(X_{.,j} - E(X_{.,j})$, where $E$ is the mathematical expectation operator. The correlation coefficient ranges from -1 to 1, where a value of $1$ implies perfect positive correlation, $-1$ implies negative correlation, and $0$ means no correlation. 
In other words, positive  correlation indicates both bands exhibit similar trends (either increasing or decreasing) while negative correlation indicates the two bands exhibits opposite trends, i.e., they diverge and hence are dissimilar. Therefore, the dissimilarity based on correlation can be derived from the similarity measures simply by deducting  the correlation coefficient from $1$. 
\begin{equation}
    D(i,j) = 1 - S(i,j).
    \label{eq:corrD}
\end{equation}

The DM thus calculated are normalized so that the values lie between $0$ and $1$. McCune and Grace \cite{mccune2002analysis} defined normalized dissimilarity measure based on the correlation coefficient in the range between $0$ and $1$ that used a scaling factor of $\frac{1}{2}$ in Eq.~\ref{eq:corrD}. However, this scaling has no effect on the normalized DM. Figure \ref{fig:DM} shows an example for the Indian Pines data set.



\subsection{Band grouping via ``visual clustering'' (iVAT and CLODD)}
After proximity measure selection and DM calculation, we perform the following steps (which are elaborated on more in subsequent subsections). Note, while the DM is a matrix, it can be displayed as an ``image'', facilitating visual analysis. If the data has any clustering tendency then the (potentially ordered) image will exhibit ``block-like'' structures. \\ 

\textbf{(Step 1)} For non-contiguous band grouping, order the bands via VAT \cite{bezdek2007visual} so that the more similar the bands are, the more closely spaced they are in the resultant DM. 

\textbf{(Step 2)} Enhance the image using iVAT by re-scaling the intensity levels so block-like structure has constant intensity levels and is more clearly distinguishable.

\textbf{(Step 3)} Identify clusters in VAT/iVAT via CLODD.\\
 
Historically, this process was referred to as visual clustering because the human was the one analyzing the DM (result of Step 1 or 2). CLODD \cite{havens2009clustering} is a clustering algorithm on VAT/iVAT. The goal of CLODD is to automate the humans visual analysis (i.e., perform clustering but do so with respect to CTA). Note, CLODD was originally applied to VAT or iVAT ordered imagery. This was done because the elements in the DM were objects and they need to be ordered for the result to make sense. However, in band grouping there is already an innate relationship typically between consecutive bands. Therefore, we introduce a new step, iVAT enhancement without ordering, to allow for contiguous band grouping. However, if the goal is non-contiguous band groups, then we perform ordering followed by enhancement then CLODD. We use  CLODD-C and CLODD-N for shorthan notations for contiguous and non-contiguous bandgrouping with CLODD respectively. Next, we discuss each step above in further detail.

\subsubsection{Band ordering via VAT}
VAT is a simple algorithm to help the human analyze if cluster tendency is present in a data set. The goal of VAT is to order the data points (bands herein) based on Prim's modified minimal single linkage (thus it leads to non-contiguous band grouping). Note, VAT was designed to be as minimalistic as possible. That is, it strives to assume little-to-nothing about a set of data and clusters therein, e.g., distance metric, compact well-separated clouds versus crazy non-linear patterns, etc. However, this is impossible as single linkage still obviously imposes some assumptions on the analysis process. However, the amount of assumptions is far less than conventional clustering techniques. Obviously, one can use a method other than single linkage if desired. First, two objects (bands) for the \textit{minimum spanning tree} (MST) are chosen such that they have maximum distance across all objects between them. Subsequently, the MST is computed on the objects. Algorithm \ref{alg:reordering_bands} is the VAT algorithm.

\begin{algorithm}
\fontsize{10pt}{20pt}\selectfont
\caption{VAT: band ordering \cite{bezdek2007visual}}
\begin{algorithmic}[1]
\State Input: $D$ ($b \times b $ DM)
\State Initialize $K = \left\{1,2, \cdots, b  \right\}$; $I = J = \emptyset$; $P = [0]^{[1 \times b]}$
\State Select $(i,j) \in \arg \ \max_{p\in K, q \in K} D(p,q)$
\State Set $P(1) = i$; $I = \left\{i \right\}$; and $J = K  \smallsetminus \left\{i \right\} $
\For {$r = 2:b $}
\State Select $(i,j) \in \arg \ \min_{p\in I, q \in J} D(p,q)$
\State Set $P(r) = j$; Replace $ I \leftarrow I \cup \left\{j \right\}$
\State $J \leftarrow J \smallsetminus \left\{j \right\}$
\EndFor
\State Obtain ordered DM, $D_R$ such that $D_R(p,q) = D(P(p), P(q))$ for $1 \le p,q \le b$
\State Output: reordered DM, $D_R$
\end{algorithmic}
\label{alg:reordering_bands}
\end{algorithm}

\subsubsection{DM visual enhancement}
In \cite{havens2012efficient}, Havens et al. proposed an \textit{improved VAT} (iVAT) that uses the graph theoretic distance to transform VAT to enhance the visualization and effectiveness of the VAT algorithm. Algorithm \ref{alg:visual_enahancement} is the enhancement step of iVAT, which is applied herein on either the raw DM, $D$, or ordered DM, $D_R$, depending on the goals of the system. 

\begin{algorithm}
\fontsize{10pt}{20pt}\selectfont
\caption{DM visual enhancement \cite{havens2012efficient}}
\begin{algorithmic}[1]
\State Input: $D_I = 
\begin{cases}
D, \quad \text{ for contiguous band grouping} \\
D_R, \quad \text{ for non-contiguous band grouping}
\end{cases}
$ 
\State Initialize $D_E = [0]^{b \times b}$
\For {$r = 2:b $}
\State Select $j = arg \min_{k \in \{ 1,2,\dots, r-1\}} D_{I}(r,k)$
\State $D_E(r,c) = D_I(r,c), c = j$
\State $D_E(r,c) = \max(D_I(r,j), D_E(j,c))$, for $c \in  \{1,2, \dots, r-1\} \smallsetminus \{j\}$
\EndFor
\State Set $D_E(c,r) = D_E(r,c), r = 2:b, c = 1,2, \cdots, r-1$
\State Output: enhanced image, $D_E$
\end{algorithmic}
\label{alg:visual_enahancement}
\end{algorithm}

\begin{figure}
\begin{center}
 \subfloat[raw DM]{\includegraphics[width=0.3\textwidth]{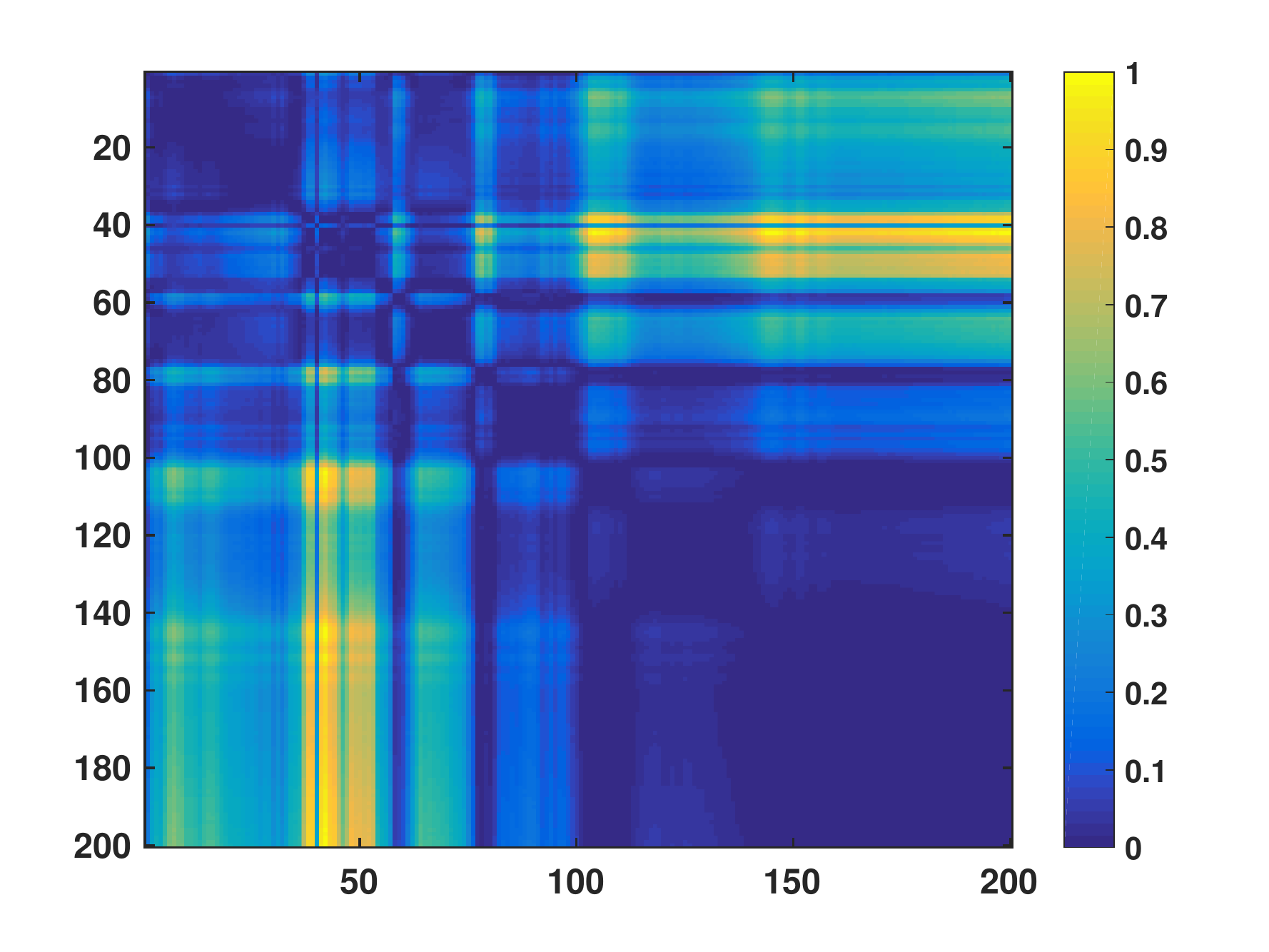}}
 \subfloat[iVAT]{\includegraphics[width=0.3\textwidth]{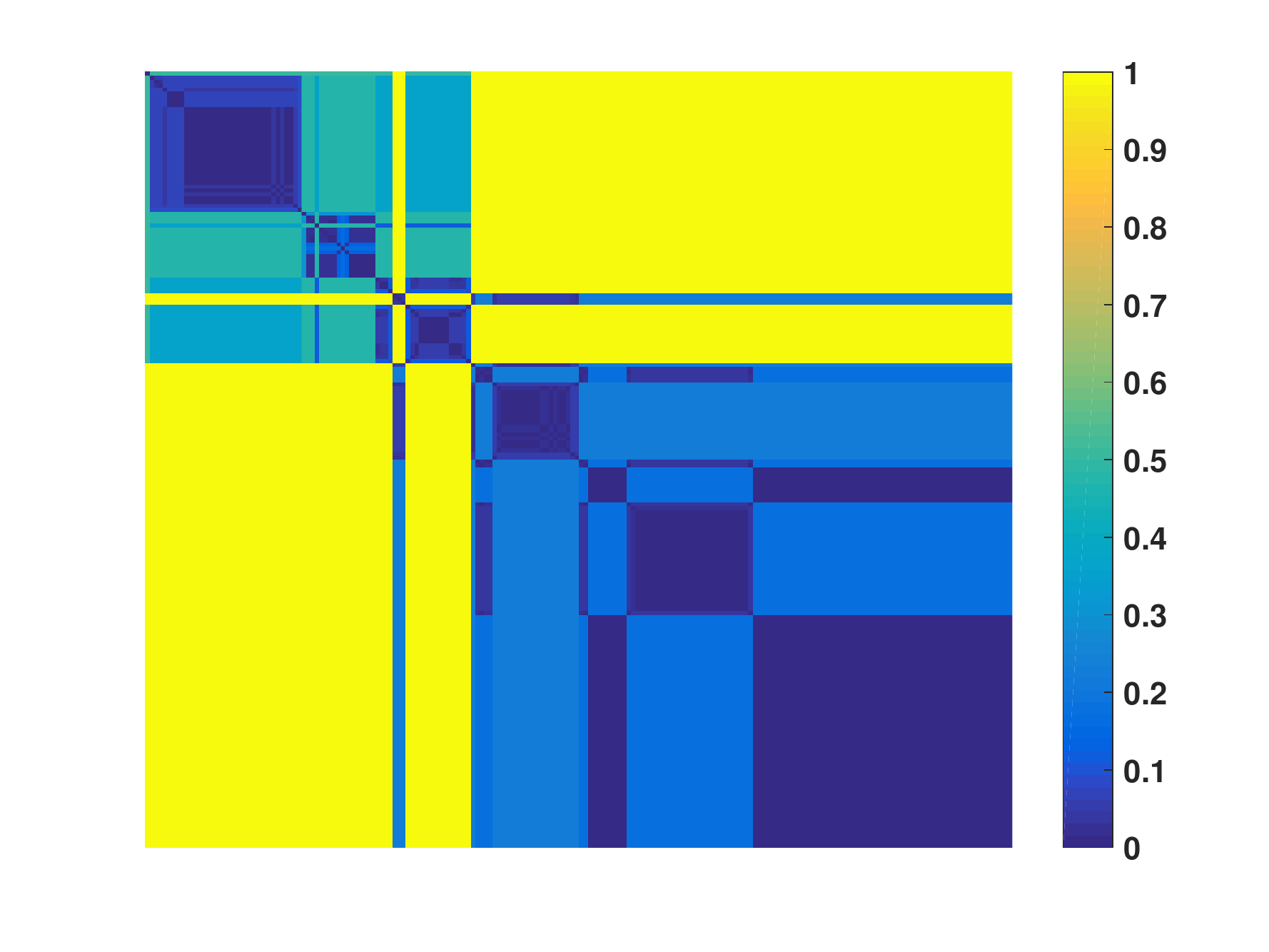}}
  \subfloat[VAT+iVAT]{\includegraphics[width=0.3\textwidth]{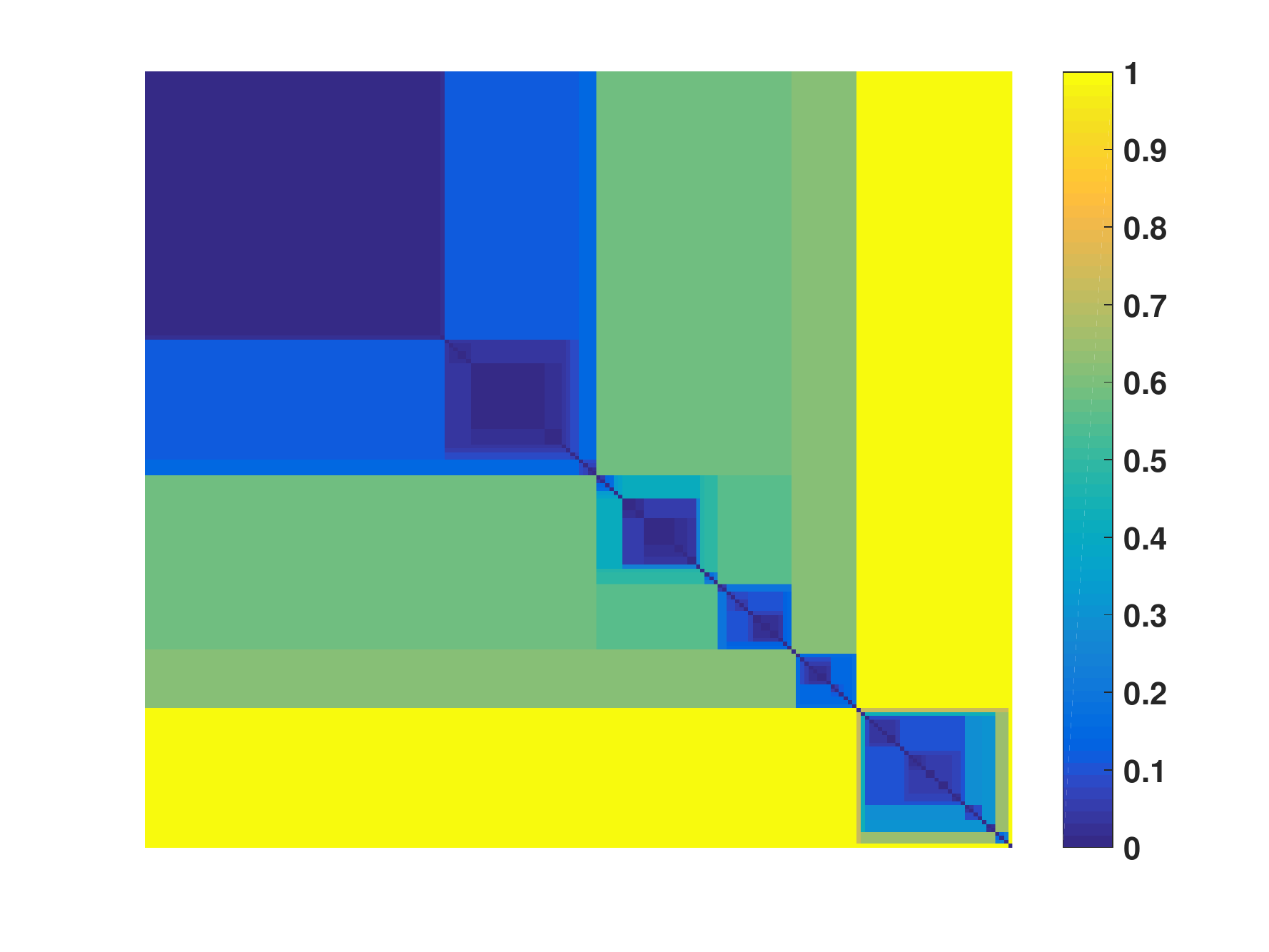}} \\
  \subfloat[raw DM]{\includegraphics[width=0.3\textwidth]{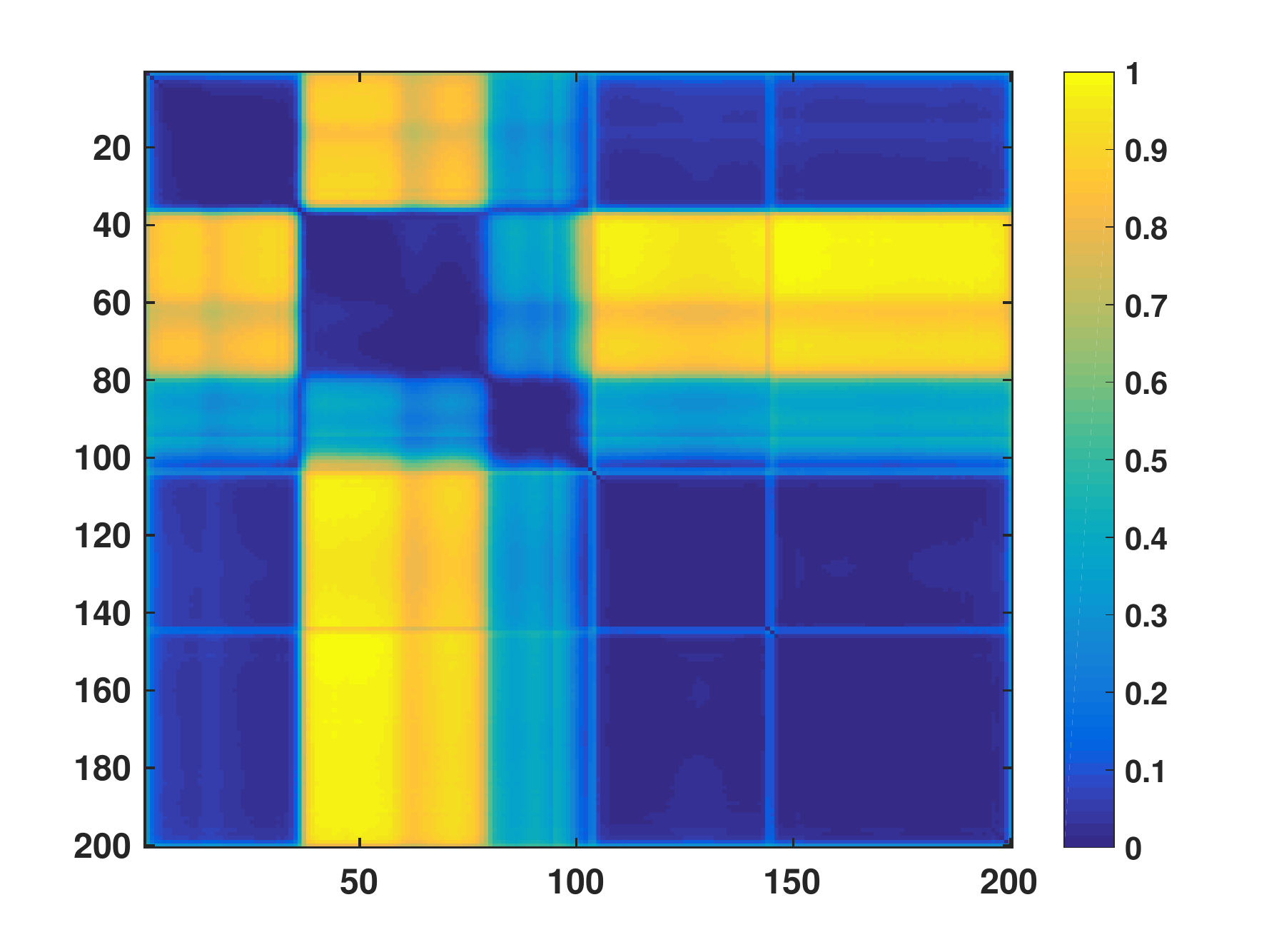}}
 \subfloat[iVAT]{\includegraphics[width=0.3\textwidth]{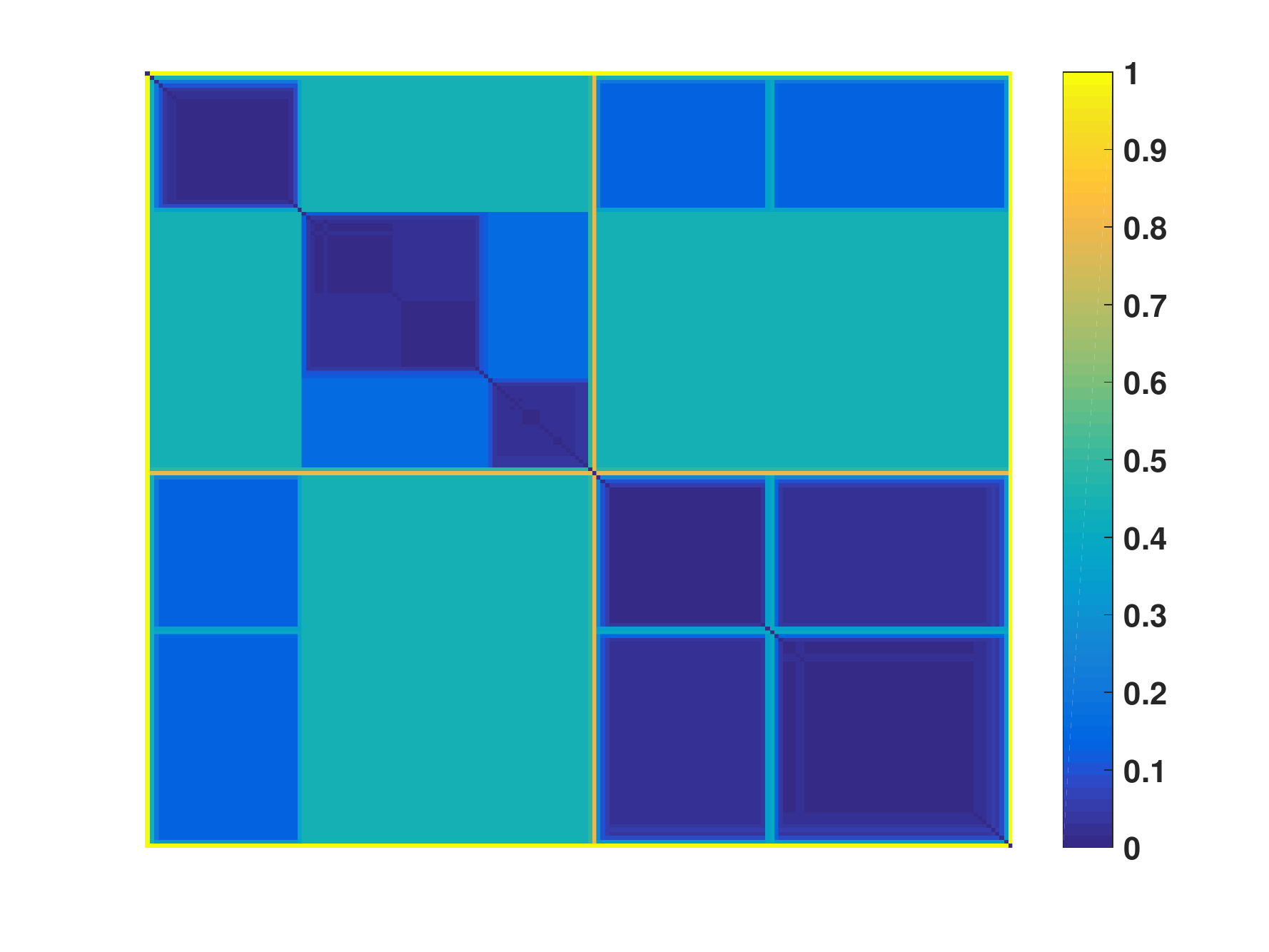}}
  \subfloat[VAT + iVAT]{\includegraphics[width=0.3\textwidth]{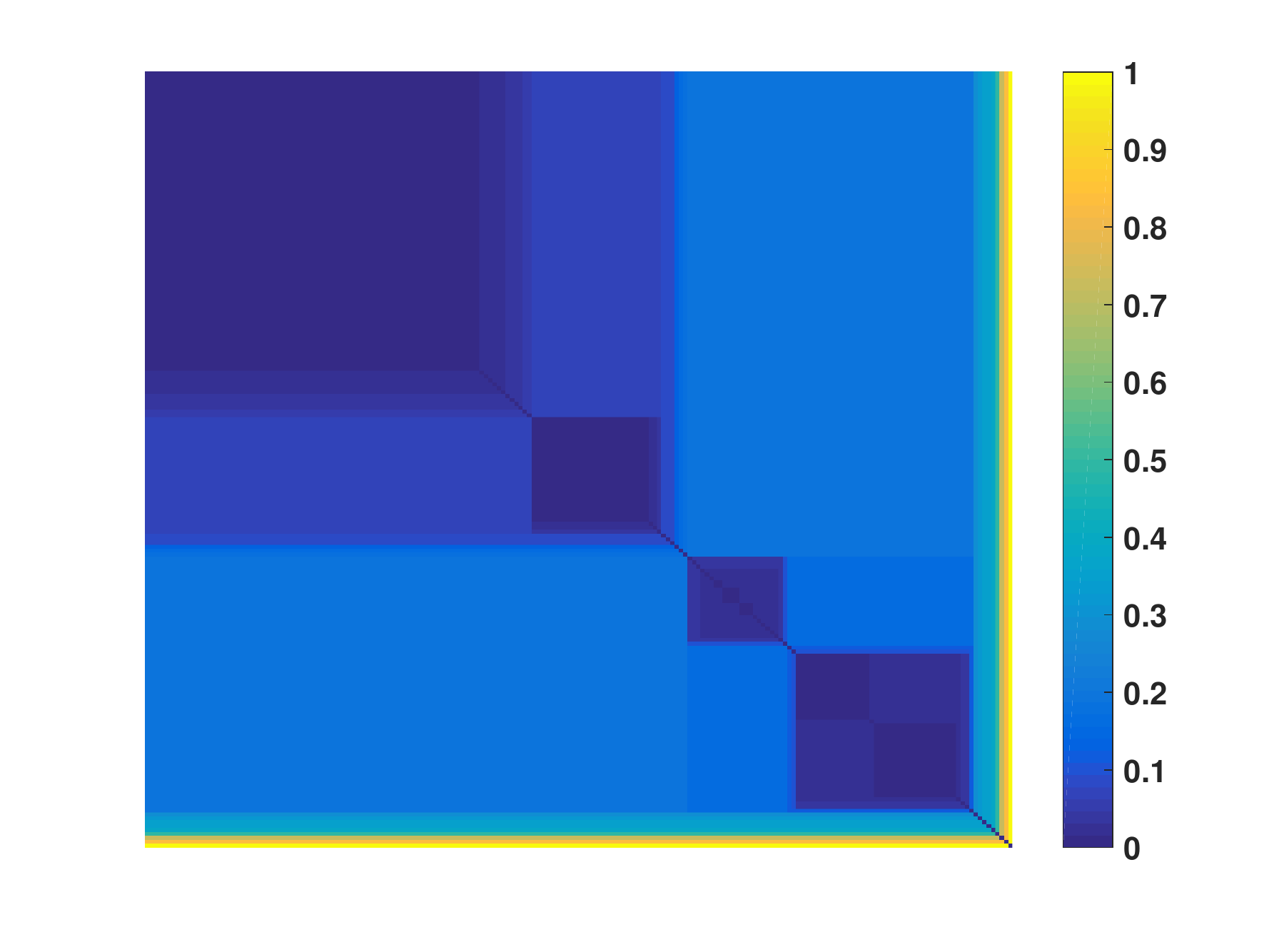}}
 \caption{DM, its iVAT enhanced image for contiguous bandgrouping, and VAT reordered + iVAT enhanced images for non-contiguous bandgrouping. Figs.~(a), (b), and (c) for the square of Euclidean and (d), (e), and (f) are for correlation.  }
 \label{fig:DM}
\end{center}
\end{figure}


Figure \ref{fig:DM} is an example of DMs for the Indian Pines data set. Block-like structure is apparent in all figures -- indicating clustering tendency for both contiguous and non-contiguous band grouping. 

\subsubsection{Clustering with CLODD}

CLODD, a clustering algorithm, is more of an image processing technique than a standard feature space clustering algorithm. CLODD looks for blockiness in a DM, VAT or iVAT output. Its goal is to find a hard partitioning (aka clusters) as dark blocks along the matrix diagonal. While searching for the partition boundaries it considers contrast between the on-diagonal dark block and off-diagonal lighter blocks known as squareness and visually apparent edges between the blocks, termed as edginess. 
 
Let $D_E$ be the iVAT input, $U$ is a $c$ partitioning, $b_i$ is the number of (contiguous) bands in cluster $i$, and $D_E(x,y)$ is the dissimilarity value corresponding to row $x$ and column $y$ in $D_E$. Squareness is

\[ E_{sq}(U;D_E) =  \frac{\sum_{i=1}^c \sum_{s \in I_i, t \notin I_i} D_E(s,t)}{\sum_{i=1}^c (b - b_i) b_i} -  \frac{\sum_{i=1}^c \sum_{s,t \in I_i, s \neq t }D_E(s,t)}{\sum_{i=1}^c (b_i^2 - b_i)}, \]
where $I_i$ is the set of indices of bands in cluster $i$.
\noindent The first part in the squareness equation is the average between dark and non-dark regions. The second is just for dark regions. Edginess is  
\begin{align*}
& E_{edge}(U;D_E) = \\
& \frac{1}{c-1} \left( \sum_{j=1}^{c-1} \frac{\sum_{i={m_j-1}}^{m_j} |D_E(i,m_j) - D_E(i,m_j + 1)|} {b_j + b_{j+1}} + 
\frac{\sum_{i={m_j + 1}}^{m_{j+1}} |D_E(i,m_j) - D_E(i,m_j + 1)|}{b_j + b_{j+1}} \right),
\end{align*}
where $m_j = \sum_{k=1}^j b_k$ and $m_0 = 1$.

The objective function has two controlling parameters: mixing coefficient, $\alpha$ to trade-off between squareness and edginess; and  $\gamma$ to impose minimum cluster size,
\begin{equation*}
E(U,D_E) = s(\min_{1\leq i \leq c} {b_i}, \gamma b) (\alpha E_{sq}(U,D_E) + (1-\alpha) E_{edge}(U,D_E)),
\end{equation*}
where s(.) is a spline function and is maximized with respect to U to obtain the optimum partition, $U^*$. Thus, in unsupervised learning, it is up to the user to provide acceptable parameters (or to possibly vary them and explore the results). After clustering, mean is used to extract a single feature from each cluster.

Now we briefly summarize our proposed method, which has the following major steps. 
    (i) Compute the DM based on different proximity measures. Herein we used the square of Euclidean and correlation.
    (ii) Reorder the bands using VAT, if non-contiguous bandgrouping is desired. Otherwise, skip this step.
    (iii) Enhance the DM image using iVAT,
    (iv) Apply CLODD to cluster the bands.
    (v) Extract a single feature from each bandgroup using mean.
    (vi) Apply diverse kernels with varying width, e.g., the square of Euclidean and correlation based RBF kernels, to generate  features in RHKS.
    (vii) Fuse the features using $l_\infty$-norm MKL, which is performed by directly summing up all the kernels and then training SVM on the resultant kernel.

\subsection{Feature Space Fusion Using $\ell_p$-Norm MKL}
\label{sec:FSMKL}

In the kernel approach, inputs are ideally projected into a high, possibly infinite, dimensional RHKS space, where the patterns for different classes are now linearly separable. The trick is that we can do this all via a ``kernel function'' in the original low(er) dimensional space and we never have to do the actual lifting. However, in reality we do not know what kernel to select and in general the choice of kernel is task specific. There is currently no straightforward way to select a kernel for a given set of data. As already mentioned, MKL provides one such path to help search for the ideal kernel by the simple concept of combining simple known (base) kernels to form custom (tailored) kernels. 

For a function to be a kernel, it need to satisfy the Mercer's kernel properties such as continuity, symmetry, and positive semi-definiteness. There are numerous kernel functions in use, e.g., \textit{radial basis function} (RBF), polynomial, etc. In this paper, we use RBF and correlation kernel. The RBF function is 
\[ k_r(\mathbf{x}_i, \mathbf{x}_j) = exp\left( - \frac{||\mathbf{x}_i^2 - \mathbf{x}_j^2||}{2 \sigma^2} \right) ,\]
where $\sigma$ is the so-called width parameter of the RBF kernel. The correlation kernel is
\[k_c(\mathbf{x}_i, \mathbf{x}_j) = exp\left(- \frac{1 - corr(\mathbf{x}_i, \mathbf{x}_j)}{2\sigma^2}\right)\]
where $corr(\mathbf{x}_i, \mathbf{x}_j)$ is the Pearson's correlation coefficient for $\mathbf{x}_i$ and $\mathbf{x}_j$. In \cite{jiang2012correlation}, the authors have shown that the correlation kernel satisfies the Mercer's kernel properties. Note, our two kernels are already more-or-less to scale by design. However, if one is using heterogeneous kernels that produce very different scales, then the zero mean and unit variance RHKS approach can be used \cite{kloft2011lp}. 

The convex sum of $M$ kernels is also a Mercer's kernel. This is because both the sum and multiplication by positive constant are \textit{positive semidefinite} (PSD) preserving operators (on $M$ different Gram matrices). The combined kernel with $\ell_p$-regularized weight $w_m$ is
\[k(\mathbf{x}_i, \mathbf{x}_j) = \sum_{m=1}^M w_m k_m(\mathbf{x}_i, \mathbf{x}_j) \]
subject to $||\mathbf{w}||_p \le 1$ and $w_m \in \mathbb{R}^+$, where $||\mathbf{w}||_p$ is the $\ell_p$-norm of $\mathbf{w}$. Though the above expression is notationally for $M$ kernels on the same set of features, it is trivially generalized to multiple features, e.g., different kernels on different subsets of features \cite{Tony2015}. Optimization-based MKL solutions, versus fixed rule or heuristic approaches, optimize (using alternating optimization typically) the weights of the kernels and the SVM criteria function. Again, we use $\ell_p$-norm MKL \cite{gu2012representative,tuia2010learning} to derive the LCS weights. However, we could use a number of other search algorithms for feature level fusion, such as MKLGL$_p$ or GAMKL$_p$, or decision-level MKL, e.g., DeFIMKL$_p$. The $\ell_p$-norm condition is more-or-less the same across a number of solvers. In general, the different approaches represent variations in search, e.g., Group Lasso (MKLGL), genetic algorithm based (GAMKLp), and non-linear decision-level fusion via DeFIMKL.  

A special case of  $\ell_p$-norm MKL is $l_{\infty}$-norm MKL, in which the combined kernel is obtained by directly summing up the individual kernel, i.e.,
\[k(\mathbf{x}_i, \mathbf{x}_j) = \sum_{m=1}^M k_m(\mathbf{x}_i, \mathbf{x}_j).\]
As noted in Section \ref{sect:intro}, this dense fusion of MKL can outperform variably weighted MKLs when combining quality features. In this case, as there is no need to perform feature selection to remove noisy kernels, performance can be maximized by just summing together all kernels. 
\section{Experimental Results}

\label{sec:experiment}
We tested our $\ell_p$-norm MKL based feature level fusion for contiguous and non-contiguous bandgrouping methods, CLODD-C and CLODD-N, on two publicly available benchmark data sets, Indian Pines and Pavia University, and their performance is compared against their counterparts, BG-Mean and Hierarchical respectively. Next, we provide a brief description of the two data sets.

\paragraph{Indian Pines\cite{PURR1947}:} The Indian Pines data set consists of $145 \times 145$ pixels with a spatial resolution of $20$ meters and $220$ spectral channels (bands). During the pre-processing of the data, we removed $20$ water absorption bands, $104-108$, $150-163$ and $220$. We considered the following $9$ classes -- Corn-notill, Corn-mintill, Grass-pasture, Grass-trees, Hay-windrowed, Soybean-notill, Soybean-mintill, Soybean-clean, and Woods, which has ($1428,830, 483, 730, 478, 972, 2455, 593, 1265$) samples respectively. 

\paragraph{Pavia University:}  The Pavia University data set has $103$ spectral bands and $610 \times 610$ pixels. The ground sampling distance is $1.3$ meters. While there are 9 classes in the image, herein we consider $7$, Gravel, Trees, Painted metal sheets, Bare Soil, Bitumen, Self-blocking bricks, and Shadows. The number of samples in these classes are $2099, 3064, 1345, 5029, 1330, 3682, \text{ and } 947$ respectively. The two remaining classes were discarded because they possessed a large number of samples ($6631$ and $18649$ respectively) for computational convenience.


\paragraph{Experimental setup:} 
The data sets were randomly partitioned into 20\% training and 80\% test sets.  The DMs were computed on the training data according to two dissimilarity measures, the squared of Euclidean and correlation. Then, we applied clustering techniques to group bands in either contiguous (CLODD-C and BG-Mean) or non-contiguous (CLODD-N and Hierarchical) manner. 
While each clustering algorithm produced a number of clustering results based on its parameter settings, we only selected the one that gave the highest accuracy.
Clustering algorithm parameters included: CLODD-C and CLODD-N scan for $\alpha$ in $[0,0.9]$ in increments of step size $0.1$, $\gamma = 3$, minimum and maximum cluster sizes for CLODD-C, CLODD-N, and Hierarchical are $5$ and $20$ respectively. For BG-Mean, bandgrouping thresholds are $(0.90, 0.95, 0.98, 0.99)$ and maximum bandgroup size $30$.
After clustering, a single feature was extracted as the mean of the bands in that group. The features were then normalized such that training features had zero mean and unit standard deviation. We used two types of Gaussian kernels, RBF and correlation, and $10$ width parameters, $(\sigma = 2^{-3}, 2^{-2}, \dots , 2^4, 2^5, 2^6)$ for each kernel type. Thus, we had the following four feature generation methods characterized by the combination of DM and kernel type used,
(M1) RBF kernel for the squared Euclidean DM,
(M2) correlation kernel for the square of Euclidean DM,
(M3) RBF kernel for the correlation DM, and
(M4) correlation kernel for the correlation DM, producing $40$ feature sets in total with $10$ for each method for different $\sigma$s. We employ one-vs-rest for multi-class classification, i.e., a model is trained for each class. During prediction,  an object is assigned a class with the highest decision value. SVM and $\ell_p$-norm MKL implementation in Shogun toolbox 4.0 \cite{sonnenburg2010shogun} have been used to conduct the experiments. 

Our goals is to compare the effectiveness of features generated from different methods, analyze how increasing $\ell_p$-norm of MKL affects the results, 
and  explore the ways to best combine features to achieve improved results. 
To this end, we investigated the following three cases, (i) SVM-based classification (no fusion), (ii) intra-method fusion, and (iii) inter-method fusion. 

\subsection{Classification using SVM}
Each method has $10$ feature sets so we ran a SVM classifier on them separately and we ranked them according to their accuracies. 
This ranking is used in in Sections \ref{subsec:intra-method} and \ref{subsec:inter-method} for intra and inter-method fusion.
We use accuracy on the validation data set to order the feature sets, which is one-half of the training set while the remaining is used for training. 

\begin{table}[htbp]
\renewcommand{\arraystretch}{1.3}
  \centering
  \caption{Indian Pines: Overall test accuracy (\%) for intra-method $\ell_p$-norm MKL}
  \resizebox{\textwidth}{!}{
    \begin{tabular}{@{}llccccccccccccc@{}}
    \toprule
          &       & \multirow{2}{*}{Top 1 kernel}     & \multicolumn{4}{c}{Top 2 kernels}                & \multicolumn{4}{c}{Top 3 kernels}      & \multicolumn{4}{c}{All kernels} \\
\cmidrule(lr){4-7}\cmidrule(lr){8-11}\cmidrule(lr){12-15}
Clustering & Method &     & $\ell_1$ & $\ell_2$ & $\ell_{100}$ & $\ell_{\infty}$ & $\ell_1$ & $\ell_2$ & $\ell_{100}$ & $\ell_{\infty}$ & $\ell_1$ & $\ell_2$ & $\ell_{100}$ & $\ell_{\infty}$ \\
    \bottomrule
    \multirow{4}{*}{CLODD-C} & sqE-rbf (M1) & 75.81 & \textbf{76.74} & \textbf{77.2}  & \textbf{77.96} & \textbf{77.98} & \textbf{76.8}  & \textbf{77.51} & \textbf{78.41} & \textbf{78.43} & \textbf{76.51} & \textbf{77.47} & \textbf{78.79} & \textbf{78.79} \\
     & sqE-corr (M2) & \textbf{75.84} & 75.83 & 76.11 & 76.6  & 76.65 & 76.44 & 76.73 & 77.14 & 77.14 & 76.45 & 76.76 & 77.38 & 77.4 \\
     & corr-rbf (M3) & 73.95 & 73.89 & 74.67 & 75.48 & 75.51 & 74.81 & 75.75 & 76.8  & 76.81 & 74.46 & 75.42 & 76.58 & 76.6 \\
     & corr-corr (M4) & 72.07 & 72.41 & 73.5  & 74.26 & 74.29 & 72.43 & 73.61 & 74.4  & 74.42 & 72.33 & 73.67 & 74.98 & 74.96 \\
     \midrule
    \multirow{4}{*}{BG-Mean} & sqE-rbf (M1) & 74.89 & 75.49 & 76.09 & 77.18 & 77.2  & 76.44 & 77.09 & 78.16 & 78.15 & 76.25 & 77.28 & 78.26 & 78.23 \\
     & sqE-corr (M2) & 74.12 & 74.17 & 74.52 & 75.05 & 75.04 & 74.65 & 75.04 & 75.54 & 75.56 & 74.6  & 75.1  & 75.81 & 75.86 \\
     & corr-rbf (M3) & 75.3  & 75.08 & 75.76 & 76.44 & 76.52 & 75.82 & 76.56 & 77.45 & 77.5  & 75.88 & 76.58 & 77.96 & 77.98 \\
     & corr-corr (M4) & 72.37 & 72.45 & 72.67 & 72.94 & 72.91 & 72.88 & 73.19 & 73.44 & 73.48 & 72.9  & 73.15 & 73.81 & 73.79 \\
     \midrule
    \multirow{4}{*}{CLODD-N} & sqE-rbf (M1) & \textbf{75.83} & 75.81 & 76.42 & 77.15 & 77.12 & 75.4  & 76.39 & 77.7  & 77.7  & 75.8  & 77.16 & 78.22 & 78.21 \\
     & sqE-corr (M2) & 74.75 & 74.63 & 75.41 & 76.18 & 76.18 & 74.86 & 75.57 & 76.33 & 76.34 & 75.06 & 75.94 & 77.1  & 77.2 \\
     & corr-rbf (M3) & 74.96 & 74.64 & 75.31 & 76.06 & 76.06 & 74.38 & 75.4  & 76.35 & 76.35 & 74.57 & 75.64 & 76.92 & 76.91 \\
     & corr-corr (M4) & 74.16 & 74.18 & 75.02 & 75.69 & 75.68 & 74.98 & 75.86 & 76.5  & 76.53 & 74.95 & 75.9  & 76.93 & 76.93 \\
     \midrule
    \multirow{4}{*}{Hierarchical} & sqE-rbf (M1) & 74.52 & 75.01 & 75.31 & 75.93 & 75.93 & 74.9  & 75.69 & 76.91 & 76.89 & 74.31 & 75.46 & 76.85 & 76.88 \\
     & sqE-corr (M2) & 71.36 & 72.27 & 72.68 & 73.26 & 73.27 & 72.59 & 72.97 & 73.64 & 73.64 & 72.61 & 73.03 & 73.91 & 73.94 \\
     & corr-rbf (M3) & 69.7  & 69.79 & 70.17 & 71.25 & 71.28 & 69.15 & 70.57 & 71.69 & 71.74 & 69.18 & 70.52 & 72    & 72.06 \\
     & corr-corr (M4) & 62.63 & 63.22 & 63.85 & 64.41 & 64.41 & 63.68 & 64.14 & 64.83 & 64.86 & 63.79 & 64.44 & 65.54 & 65.53 \\
    \bottomrule
    \end{tabular}%
    }
  \label{tab:intra-indian-pines}%
\end{table}%

\begin{table}[htbp]
\renewcommand{\arraystretch}{1.3}
  \centering
  \caption{Pavia University: Overall accuracy (\%) for intra-method fusion}
  \resizebox{\textwidth}{!}{
 \begin{tabular}{@{}llccccccccccccc@{}}
    \toprule
          &       & \multirow{2}{*}{Top 1 kernel}     & \multicolumn{4}{c}{Top 2 kernels}                & \multicolumn{4}{c}{Top 3 kernels}      & \multicolumn{4}{c}{All kernels} \\
\cmidrule(lr){4-7}\cmidrule(lr){8-11}\cmidrule(lr){12-15}
Clustering & Method &     & $\ell_1$ & $\ell_2$ & $\ell_{100}$ & $\ell_{\infty}$ & $\ell_1$ & $\ell_2$ & $\ell_{100}$ & $\ell_{\infty}$ & $\ell_1$ & $\ell_2$ & $\ell_{100}$ & $\ell_{\infty}$ \\
    \bottomrule
    \multirow{4}{*}{CLODD-C} & sqE-rbf (M1) & 91.91 & 92.61 & 93.05 & 93.35 & 93.36 & 93.54 & 94.03 & 94.58 & 94.61 & 93.86 & 94.58 & \textbf{94.97} & \textbf{94.96} \\
     & sqE-corr (M2) & 90.39 & 90.63 & 90.73 & 91.15 & 91.17 & 91.8  & 91.95 & 92.25 & 92.25 & 92.67 & 92.84 & 93.02 & 93.03 \\
     & corr-rbf (M3) & 91.87 & 92.3  & 92.81 & 93.17 & 93.18 & 93.23 & 93.75 & 94.19 & 94.21 & 93.78 & 94.36 & 94.86 & 94.86 \\
     & corr-corr (M4) & 89.75 & 90.53 & 90.7  & 91.01 & 91    & 91.75 & 91.83 & 92.1  & 92.09 & 92.31 & 92.4  & 92.67 & 92.67 \\
     \midrule
    \multirow{4}{*}{BG-Mean} & sqE-rbf (M1)     & 90.45 & 90.64 & 90.66 & 90.8  & 90.82 & 91.81 & 91.84 & 92.01 & 92.01 & 92.3  & 92.48 & 92.54 & 92.55 \\
     & sqE-corr (M2)     & 85.51 & 85.9  & 85.93 & 85.98 & 85.98 & 86.07 & 86.11 & 86.12 & 86.12 & 86.32 & 86.36 & 86.39 & 86.38 \\
     & corr-rbf (M3)     & 91.07 & 92.21 & 92.34 & 92.48 & 92.48 & 92.33 & 92.39 & 92.6  & 92.61 & 92.85 & 93.01 & 93.01 & 93 \\
     & corr-corr (M4)     & 85.75 & 86.11 & 86.14 & 86.23 & 86.25 & 86.55 & 86.54 & 86.67 & 86.67 & 86.7  & 86.67 & 86.83 & 86.83 \\
     \midrule
    \multirow{4}{*}{CLODD-N} & sqE-rbf (M1)     & \textbf{94.19} & \textbf{94.27} & \textbf{94.46} & \textbf{94.63} & \textbf{94.63} & \textbf{94.31} & \textbf{94.6}  & \textbf{94.83} & \textbf{94.81} & \textbf{94.46} & \textbf{94.74} & \textbf{94.96} & \textbf{94.95} \\
     & sqE-corr (M2)     & 86.67 & 89.07 & 89.22 & 89.46 & 89.47 & 90.65 & 90.7  & 90.68 & 90.68 & 91.1  & 91.25 & 91.62 & 91.65 \\
     & corr-rbf (M3)     & 92.85 & 93.15 & 93.44 & 93.72 & 93.73 & 93.83 & 94.09 & 94.31 & 94.31 & 93.54 & 93.93 & 94.16 & 94.17 \\
     & corr-corr (M4)     & 86.21 & 88.42 & 88.48 & 88.78 & 88.77 & 90.6  & 90.69 & 90.76 & 90.78 & 91.02 & 91.17 & 91.5  & 91.51 \\
     \midrule
    \multirow{4}{*}{Hierarchical} & sqE-rbf (M1)     & 91.65 & 92.25 & 92.42 & 92.7  & 92.7  & 92.39 & 92.65 & 93.02 & 93.04 & 93.62 & 93.81 & 94.12 & 94.14 \\
     & sqE-corr (M2)     & 89    & 89.14 & 89.3  & 89.64 & 89.65 & 89.46 & 89.63 & 89.92 & 89.93 & 89.6  & 89.74 & 90.08 & 90.08 \\
     & corr-rbf (M3)     & 91.66 & 91.7  & 92.17 & 92.46 & 92.46 & 93.02 & 93.45 & 93.84 & 93.85 & 93.56 & 93.9  & 94.43 & 94.43 \\
     & corr-corr (M4)     & 89    & 89.21 & 89.45 & 90    & 90.02 & 90.11 & 90.38 & 90.78 & 90.8  & 90.74 & 90.94 & 91.28 & 91.3 \\
     \bottomrule
    \end{tabular}%
    }
  \label{tab:intra-pavia}%
\end{table}%

The third columns in Tables \ref{tab:intra-indian-pines} and  \ref{tab:intra-pavia} report the best overall test accuracies for using a SVM classifier based on the Indian Pines and Pavia University datasets respectively. As we can see, CLODD-C leads in contiguous bandgroping and CLODD-N in non-contiguous bandgrouping.  Overall, both CLODD-C and CLODD-N are the best for Indian Pines whereas CLODD-N is the best for Pavia University. Based on these results, we can say that CLODD  is better than all other clustering algorithms in comparison (whether contiguous or non-contiguous) and non-contiguous variant of CLODD has the potential to have better performance in some cases, but at the expense of a more complex (and possibly non-realizable) physical sensor.

To no surprise, there is no clear winner when it comes to feature generation methods. 
For example, sqE-corr, corr-rbf, and sqE-rbf provide the best results for CLODD-C, BG-Mean, and CLODD-N respectively on Indian Pines dataset. In general, corr-rbf is  good for BG-Mean and sqE-rbf for the rest while corr-corr always comes last. 

\subsection{Intra-method fusion}
\label{subsec:intra-method}
In this experiment, $\ell_p$-norm MKL is used to fuse sets of features arising from a method, particularly the top-2, top-3, and all (10) features ranked by SVM classification in the previous subsection. We use three $\ell_p$-norms, $\ell_{1.01}, \ell_{2}, \ell_{100}, \ell_{\infty}$-norm MKL to investigate how regularization influences the results and to identify the best $\ell_p$-norm, computationally and accuracy-wise, for the problem at hand. Note that in the implementation of $\ell_p$-norm, $p$ cannot be set to $1$ so we used $1.01$ to get $\ell_1$ equivalent results. 
Based on the results shown in Tables \ref{tab:intra-indian-pines} and  \ref{tab:intra-pavia}, we can make few observations. First, there is clearly a trend that increasing $\ell_p$-norm  improves the classification results, however, its impact diminishes as $p$ grows.   Depending on the quality and diversity of the features, highest accuracy may be achieved at a lower $p$ well before $\ell_{\infty}$-norm.  
We would like to point out here that even in such a case, $\ell_{\infty}$ is the best option to consider primarily because $\ell_{\infty}$ is efficient and does not hurt results when the the features are good. Furthermore, parameter search to find the exact $p$, at which the accuracy saturates, could add huge computational burden to the process. Second, increasing the number of kernels also improves the result but has diminishing returns. For example, fusing top two kernels in M1 increases the accuracy by $1.17\%$ while adding $7$ more kernels (all features) improves the results by only $0.81\%$ (third row in Table \ref{tab:intra-indian-pines}). Third, method M1 (sqE-rbf) appears to have better results for $\ell_p$-norm MKL based fusion/fusing multiple feature sets even though a different method might give be better for SVM, e.g., M3 (corr-rbf) for BG-mean and M2 (sqE-corr) for CLODD-C on Indian Pines dataset.

\subsection{Inter-method fusion}
\label{subsec:inter-method}
In this subsection, we perform cross-method fusion---i.e., aggregate feature sets across multiple methods---for the following five scenarios: (i) all methods, (ii) M1 and M2, (iii) M3 and M4, (iv) M1 and M3, and (v) M2 and M4. Tables \ref{tab:indian-pines-inter-method} and \ref{tab:pavia-inter-method} show the fusion results of top-1, top-2, and top-3 features from each method for Indian Pines and Pavia University respectively. We did not consider all features because the gain in accuracy from top-2 to top-3 sets is quite small, 0.23\% for Indian Pines and 0.24\% for Pavia University. 

\begin{table}[htbp]
\renewcommand{\arraystretch}{1.3}
  \centering
  \caption{Indian Pines: Overall Accuracy for $\ell_p$-norm MKL based fusion }
    \resizebox{\textwidth}{!}{
    \begin{tabular}{@{}llcccccccccccc@{}}
    \toprule
          &       & \multicolumn{4}{c}{Top 1 kernel}      & \multicolumn{4}{c}{Top 2 kernels}    & \multicolumn{4}{c}{Top 3 kernels} \\
          \cmidrule(lr){3-6}\cmidrule(lr){7-10}\cmidrule(lr){11-14}

    Clustering & Methods & $\ell_1$ & $\ell_2$ & $\ell_{100}$ & $\ell_{\infty}$ & $\ell_1$ & $\ell_2$ & $\ell_{100}$ & $\ell_{\infty}$ & $\ell_1$ & $\ell_2$ & $\ell_{100}$ & $\ell_{\infty}$ \\
    \midrule
    \multirow{5}{*}{CLODD-C} & M1,M2,M3,M4   & 78.17 & \textbf{79.79} & \textbf{80.68} & \textbf{80.66} & 78.23 & \textbf{80.06} & \textbf{81.06} & \textbf{81.06} & 78.47 & \textbf{80.26} & \textbf{81.29} & \textbf{81.29} \\
     & M1,M2   & \textbf{78.21} & 79.15 & 79.78 & 79.81 & \textbf{78.7}  & 79.77 & 80.89 & 80.91 & \textbf{78.5}  & 79.77 & 81.09 & 81.08 \\
     & M3,M4   & 75.3  & 76.34 & 77.11 & 77.12 & 75.4  & 77.2  & 78.16 & 78.13 & 76.06 & 77.52 & 78.46 & 78.47 \\
     & M1,M3   & 76.1  & 76.93 & 77.7  & 77.74 & 76.29 & 77.55 & 79.13 & 79.1  & 76.58 & 78.05 & 79.37 & 79.37 \\
     & M2,M4   & 75.93 & 76.67 & 77.33 & 77.34 & 75.71 & 77.21 & 78.46 & 78.5  & 75.76 & 77.3  & 78.51 & 78.54 \\
     \midrule
    \multirow{5}{*}{BG-Mean} & M1,M2,M3,M4   & 77.8  & 79.31 & 80.1  & 80.15 & 78.02 & 79.92 & 80.61 & 80.6  & 78.58 & 79.98 & 80.6  & 80.66 \\
     & M1,M2   & 77.51 & 78.58 & 79.44 & 79.48 & 77.93 & 79.25 & 80.44 & 80.47 & 78.38 & 79.51 & 80.72 & 80.72 \\
     & M3,M4   & 76.48 & 77.05 & 77.7  & 77.69 & 76.58 & 77.4  & 78.91 & 78.91 & 77.16 & 78.05 & 79.39 & 79.42 \\
     & M1,M3   & 75.54 & 76.42 & 77.39 & 77.41 & 75.72 & 77.12 & 78.8  & 78.81 & 76.46 & 77.8  & 78.91 & 78.89 \\
     & M2,M4   & 74.72 & 75.7  & 76.45 & 76.47 & 75.11 & 76.12 & 77.2  & 77.23 & 75.47 & 76.29 & 77.2  & 77.24 \\
     \midrule
    \multirow{5}{*}{CLODD-N} & M1,M2,M3,M4   & 77.91 & 79.02 & 79.67 & 79.66 & 77.97 & 79.63 & 80.56 & 80.59 & 77.99 & 79.71 & 80.44 & 80.45 \\
     & M1,M2   & 77.57 & 78.32 & 78.68 & 78.7  & 77.47 & 78.92 & 79.63 & 79.61 & 77.1  & 78.64 & 80.06 & 80.09 \\
     & M3,M4   & 76.87 & 77.47 & 78.21 & 78.21 & 76.64 & 78.13 & 79.51 & 79.5  & 76.98 & 78.57 & 79.54 & 79.54 \\
     & M1,M3   & 76    & 76.48 & 76.94 & 76.95 & 75.78 & 76.65 & 77.76 & 77.75 & 75.41 & 76.47 & 77.91 & 77.88 \\
     & M2,M4   & 75.92 & 76.63 & 77.49 & 77.49 & 75.84 & 77.17 & 78.4  & 78.37 & 76.53 & 77.73 & 78.98 & 78.99 \\
     \midrule
    \multirow{5}{*}{Hierarchical} & M1,M2,M3,M4   & 76.7  & 77.33 & 77.99 & 78    & 76.64 & 77.69 & 78.62 & 78.63 & 76.42 & 78.06 & 79.01 & 79.05 \\
     & M1,M2   & 76.67 & 77.08 & 77.73 & 77.75 & 76.58 & 77.29 & 78.55 & 78.54 & 76.4  & 78.22 & 79.52 & 79.5 \\
     & M3,M4   & 70.44 & 70.93 & 71.5  & 71.5  & 70.5  & 71.69 & 72.42 & 72.43 & 70.58 & 71.89 & 72.96 & 72.97 \\
     & M1,M3   & 75.01 & 75.19 & 75.58 & 75.58 & 75.16 & 75.71 & 76.76 & 76.82 & 74.83 & 75.9  & 77.03 & 77.09 \\
     & M2,M4   & 72.31 & 72.72 & 73.18 & 73.23 & 73.27 & 73.84 & 74.94 & 74.94 & 73.01 & 73.88 & 75.24 & 75.27 \\
     \bottomrule
    \end{tabular}%
    }
  \label{tab:indian-pines-inter-method}%
\end{table}%

\begin{table}[htbp]
\renewcommand{\arraystretch}{1.3}
  \centering
  \caption{Pavia University: Overall Accuracy for $\ell_p$-norm MKL based fusion}
  \resizebox{\textwidth}{!}{
   \begin{tabular}{@{}llcccccccccccc@{}}
    \toprule
          &       & \multicolumn{4}{c}{Top 1 kernel}      & \multicolumn{4}{c}{Top 2 kernels}    & \multicolumn{4}{c}{Top 3 kernels} \\
          \cmidrule(lr){3-6}\cmidrule(lr){7-10}\cmidrule(lr){11-14}

    Clustering & Methods & $\ell_1$ & $\ell_2$ & $\ell_{100}$ & $\ell_{\infty}$ & $\ell_1$ & $\ell_2$ & $\ell_{100}$ & $\ell_{\infty}$ & $\ell_1$ & $\ell_2$ & $\ell_{100}$ & $\ell_{\infty}$ \\
    \midrule
    \multirow{5}{*}{CLODD-C} & M1,M2,M3,M4   & 93.35 & 94.06 & 94.77 & 94.8  & 93.81 & 94.94 & \textbf{95.55} & \textbf{95.55} & 94.97 & \textbf{95.52} & \textbf{95.78} & \textbf{95.79} \\
     & M1,M2   & 93.06 & 93.61 & 94.01 & 94.01 & 93.55 & 94.4  & 95.08 & 95.11 & 94.83 & 95.35 & 95.66 & 95.66 \\
     & M3,M4   & 92.8  & 93.35 & 93.84 & 93.83 & 93.68 & 94.48 & 95.02 & 95.03 & 94.9  & 95.21 & 95.34 & 95.36 \\
     & M1,M3   & 92.03 & 92.28 & 92.73 & 92.73 & 92.63 & 93.38 & 94    & 94.02 & 93.59 & 94.46 & 95.2  & 95.21 \\
     & M2,M4   & 90.53 & 90.85 & 91.53 & 91.54 & 91.02 & 91.6  & 92.22 & 92.22 & 92    & 92.4  & 92.73 & 92.74 \\
     \midrule
    \multirow{5}{*}{BG-Mean} & M1,M2,M3,M4   & 91.18 & 91.61 & 92.3  & 92.33 & 92.37 & 92.58 & 93.05 & 93.06 & 92.58 & 92.98 & 93.3  & 93.32 \\
     & M1,M2   & 90.45 & 90.54 & 90.68 & 90.68 & 90.68 & 90.86 & 91    & 91    & 91.63 & 91.89 & 92    & 92.01 \\
     & M3,M4   & 91.28 & 91.38 & 91.69 & 91.7  & 92.36 & 92.56 & 92.89 & 92.89 & 92.53 & 92.88 & 93.16 & 93.19 \\
     & M1,M3   & 91.07 & 91.36 & 91.67 & 91.68 & 92.24 & 92.38 & 92.64 & 92.64 & 92.38 & 92.68 & 92.97 & 92.97 \\
     & M2,M4   & 86.72 & 86.83 & 87.03 & 87.03 & 87.05 & 87.33 & 87.55 & 87.54 & 87.37 & 87.65 & 87.95 & 87.95 \\
     \midrule
    \multirow{5}{*}{CLODD-N} & M1,M2,M3,M4   & \textbf{94.78} & \textbf{95.05} & \textbf{95.33} & \textbf{95.33} & \textbf{94.87} & \textbf{95.18} & 95.43 & 95.44 & 94.86 & 95.25 & 95.63 & 95.62 \\
     & M1,M2   & 94.76 & 94.89 & 95.06 & 95.06 & 94.85 & 95.1  & 95.47 & 95.47 & \textbf{95.03} & 95.25 & 95.53 & 95.55 \\
     & M3,M4   & 93.33 & 93.85 & 94.28 & 94.28 & 94.03 & 94.54 & 94.92 & 94.93 & 94.59 & 94.83 & 95.05 & 95.05 \\
     & M1,M3   & 94.31 & 94.48 & 94.62 & 94.64 & 94.26 & 94.63 & 94.81 & 94.81 & 94.11 & 94.68 & 94.92 & 94.91 \\
     & M2,M4   & 86.9  & 86.77 & 86.94 & 86.96 & 89.32 & 89.57 & 89.57 & 89.59 & 91.26 & 91.13 & 90.66 & 90.63 \\
     \midrule
    \multirow{5}{*}{Hierarchical} & M1,M2,M3,M4   & 92.8  & 93.59 & 94.21 & 94.21 & 92.84 & 93.83 & 94.58 & 94.6  & 93.92 & 94.58 & 94.92 & 94.94 \\
     & M1,M2   & 92.62 & 93.03 & 93.33 & 93.35 & 92.9  & 93.28 & 93.74 & 93.77 & 92.78 & 93.46 & 94.06 & 94.07 \\
     & M3,M4   & 92.3  & 92.76 & 93.08 & 93.08 & 92.27 & 93.09 & 93.84 & 93.87 & 93.62 & 94.37 & 94.79 & 94.79 \\
     & M1,M3   & 91.98 & 92.65 & 93.15 & 93.15 & 92.18 & 92.98 & 93.68 & 93.68 & 93.37 & 93.93 & 94.39 & 94.41 \\
     & M2,M4   & 89.29 & 89.68 & 90.18 & 90.19 & 89.53 & 90.14 & 90.78 & 90.78 & 90.58 & 91.19 & 91.62 & 91.61 \\
    \bottomrule
    \end{tabular}%
    }
  \label{tab:pavia-inter-method}%
\end{table}%


The results tells us that fusion across methods is better than within-method fusion. For example, cross-method fusion of M1 and M2 with only two features yield an accuracy of  $79.81\%$ (row 4 in Table \ref{tab:indian-pines-inter-method}), which is higher than the best accuracy for any method in intra-MKL fusion ($78.79\%$ accuracy with $10$ feature sets). Since proximity measures and/or kernel type varies from method to method, the features produced by different methods are diverse and thus cause an uptick in the performance when fused. On the other hand, the feature sets within a method only differs by width parameter, $\sigma$, so resulting in little gain in compared to inter-method fusion.

While CLODD-N performs better on Pavia University for intra-method and inter-method fusion of top-1, and some cases of top-2 feature sets, CLODD-C takes over CLODD-N at the fusion of top-3 sets. 
This could be due to CLODD-C features possibly having more diversity than CLODD-N. At the moment, our method is an end-to-end solution. It is about performance and not understandability. In future work, we will consider interesting ways to help us open the hood and understand what performance gain is associated with, e.g., more diverse features.

\section{Conclusion}
In this article, we explored a dimensionality reduction technique using a visual clustering algorithm, CLODD, which in conjunction with VAT and iVAT can cluster bands in contiguous and non-contiguous manner. We showed through our experiments that CLODD-C outperform BG-Mean in contiguous bandgrouping and CLODD-N outperforms Hierarchical in non-contiguous bandgrouping. 
We proposed an end-to-end fusion technique based on CLODD that generate features through different types of proximity measures and Gausssian kernels, which are than fused using $\ell_{\infty}$-norm MKL. The experiment results show that this cross-method feature level fusion outperforms than any individual technique and $\infty$-norm MKL outperforms SVM and other MKLs. 
We advocate the use of $\ell_{\infty}$-norm MKL for feature level fusion instead of commonly used  $\ell_1$ and $\ell_2$-norm MKL in hyperspectral image processing, which is computationally efficient and has the best result.

While we demonstrated our proposed method with only two proximity measures and two kernels, more proximity measure and kernels could be explored to further enhance the results.
In the future, we will also explore deep learning algorithms to train on multiple sets of features extracted using proposed approach. For inter-method fusion, we picked the same number of features (top-2 or top-3) from each method, however, different combination of features could maximize the performance. Therefore, we will focus on developing more sophisticated way to rank and select feature sets across different feature extractions methods.



\bibliography{refs}   
\bibliographystyle{spiejour}   




\vspace{2ex}\noindent\textbf{Muhammad Aminul Islam} is a Research Assistant Professor in the Center for Geospatial Intelligence at the University of Missouri. His research interests include data/information fusion, machine learning, deep learning, autonomous driving, and hyperspectral image processing.

\vspace{1ex}

\vspace{2ex}\noindent\textbf{John E. Ball} is an Assistant Professor of Electrical and Computer Engineering at Mississippi State University (MSU), USA. He received his Ph.D. degree from MSU in 2007. He is a Co-Director of the Sensor Analysis and Intelligence Laboratory (SAIL) at MSU. He has authored 45+ articles, and 22 technical tutorials and reports. His research interests are: deep learning, remote sensing, signal processing, and image processing. He has approximately \$6M in research funding to date. He is an Associate Editor for the SPIE Journal of Applied Remote Sensing. 

\vspace{1ex}
\vspace{2ex}\noindent\textbf{Derek T. Anderson} is an Associate Professor in the Electrical Engineering and Computer Science (EECS) department at the University of Missouri-Columbia (MU), an Intermittent Faculty Member with the Naval Research Laboratory, and Director of the Mizzou INformation and Data FUsion Laboratory (MINDFUL). Prior to joining MU, he was an Associate Professor and Robert D. Guyton Chair in Electrical and Computer Engineering at Mississippi State University and Co-Director of the Sensor Analysis and Intelligence Laboratory (SAIL). His research is data/information fusion for machine learning and automated decision making in signal/image processing and computer vision. He has published 120+ articles, he is an Associate Editor for the IEEE Transactions on Fuzzy Systems, Program Co-Chair of FUZZ--IEEE 2019, and a Program Committee member for detection and sensing of mines, explosive objects, and obscured targets at SPIE.


\listoffigures
\listoftables

\end{spacing}
\end{document}